%% file: RJwrapper.tex
\documentclass[a4paper]{report}
\usepackage[utf8]{inputenc}
\usepackage[T1]{fontenc}
\usepackage{RJournal}
\usepackage{amsmath,amssymb,array}
\usepackage{booktabs}

\usepackage{pdflscape}
\usepackage{float}
\newcommand{\blandscape}{\begin{landscape}}
\newcommand{\elandscape}{\end{landscape}}

\DeclareMathOperator*{\argmin}{argmin}

\newcommand{\1}{\ensuremath{\mathbf{1}}}

\newcommand{\T}{\ensuremath{\widetilde{T}}}
\newcommand{\X}{\ensuremath{{W}}}

\newcommand{\ax}{\ensuremath{\mid a,\,{w}}}
\newcommand{\aX}{\ensuremath{\mid A = a,\,{W}}}
\newcommand{\AX}{\ensuremath{\mid A,\,{W}}}
\newcommand{\x}{\ensuremath{{w}}}
\newcommand{\trt}{\ensuremath{\pi^*}}

\newcommand{\lj}{\ensuremath{l}}
\newcommand{\jj}{\ensuremath{j}}

\newcommand{\g}{\ensuremath{\pi}}
\renewcommand{\L}{\ensuremath{W}}
\renewcommand{\l}{\ensuremath{w}}
\newcommand{\tDelta}{\ensuremath{\widetilde{\Delta}}}

\begin{document}

\sectionhead{Contributed research article}
\volume{XX}
\volnumber{YY}
\year{20ZZ}
\month{AAAA}

\begin{article}
  \input{concrete}
\end{article}

\end{document}

%% file: concrete.tex
\title{\pkg{concrete}: Targeted Estimation of Survival and Competing Risks Estimands in Continous Time}
\author{by David Chen, Helene C. W. Rytgaard, Edwin C. H. Fong, Jens M. Tarp, Maya L. Petersen, Mark J. van der Laan, Thomas A. Gerds}

\maketitle

\abstract{
This article introduces the R package concrete, which implements a recently developed targeted maximum likelihood estimator (TMLE) for the cause-specific absolute risks of time-to-event outcomes measured in continuous time. Cross-validated Super Learner machine learning ensembles are used to estimate propensity scores and conditional cause-specific hazards, which are then targeted to produce robust and efficient plug-in estimates of the effects of static or dynamic interventions on a binary treatment given at baseline quantified as risk differences or risk ratios. Influence curve-based asymptotic inference is provided for TMLE estimates and simultaneous confidence bands can be computed for target estimands spanning multiple times or events. In this paper we review the one-step continuous-time TMLE methodology as it is situated in an overarching causal inference workflow, describe its implementation, and demonstrate the use of the package on the PBC dataset.
}

\section{Introduction}
\label{intro}
In biomedical applications evaluating treatment effects on time-to-event outcomes, study subjects are often susceptible to competing risks such as all-cause mortality. In recent decades, several competing risk methods have been developed; including the Fine-Gray subdistributions model \citep{fine_proportional_1999}, cause-specific Cox regression \citep{benichou_estimates_1990}, pseudovalue \citep{klein_regression_2005}, and direct binomial \citep{scheike_predicting_2008, gerds_absolute_2012} regressions; and authors have consistently cautioned against the use of standard survival estimands for causal questions involving competing risks. Nevertheless, reviews of clinical literature \citep{koller_competing_2012, austin_accounting_2017} found that most trials still fail to adequately address the effect of potential competing risks in their studies. Meanwhile, formal causal inference frameworks \citep{rubin_estimating_1974, pearl_causal_2016} gained recognition for their utility in translating clinical questions into statistical analyses and the targeted maximum likelihood estimation (TMLE) \citep{laan_targeted_2006, laan_targeted_2011, laan_targeted_2018} methodology was developed from the estimating equation and one-step estimator lineage of constructing semi-parametric efficient estimators through solving efficient influence curve equations. The targeted learning roadmap \citep{petersen_causal_2014} combines these developments into a cohesive causal inference workflow and provides a structured way to think about statistical decisions. In this paper we apply the targeted learning roadmap to an analysis of time-to-event outcomes and demonstrate the R package \CRANpkg{concrete}, which implements a recently developed continuous-time TMLE targeting cause-specific absolute risks \citep{rytgaard_one-step_2021, rytgaard_targeted_2022, rytgaard_estimation_2023}.

Given identification and regularity assumptions, \pkg{concrete} can be used to efficiently estimate the treatment effect of interventions given at baseline. In short, the implemented one-step TMLE procedure consists of three stages: 1) an initial estimation of nuisance parameters, 2) a targeted update of the initial estimators to solve the estimating equation corresponding to the target statistical estimand's efficient influence curve (EIC), and 3) a plug-in of the updated estimators into the original parameter mapping to produce a substitution estimator of the target estimand. 

In \pkg{concrete} the initial nuisance parameter estimation is performed using Super Learning, a cross-validated machine-learning ensemble algorithm with asymptotic oracle guarantees \citep{laan_unified_2003, laan_super_2007, polley_superlearner_2021}, as flexible machine-learning approaches such as Super Learners with robust candidate libraries and appropriate loss functions often give users the best chance of achieving the convergence rates needed for TMLE's asymptotic properties. The subsequent targeted update is based on semi-parametric efficiency theory in that efficient regular and asymptotically linear (RAL) estimators must have influence curves equal to the efficient influence curve (EIC) of the target statistical estimand, see e.g. \citep{bickel_efficient_1998, laan_targeted_2011, laan_targeted_2018, kennedy_semiparametric_2016}. In TMLE, initial nuisance parameter estimates are updated to solve the estimating equation corresponding to the target EIC, thus recovering normal asymptotic inference (given that initial estimators converge at adequate rates) while leveraging flexible machine-learning algorithms for initial estimation. In Section \ref{cv} we outline how Super Learner is used to estimate nuisance parameters in \pkg{concrete}; more detailed guidance on how to best specify Super Learner estimators is provided in e.g., \citep{phillips_practical_2023, dudoit_asymptotics_2005, vaart_oracle_2006}. Section \ref{EIC} outlines the subsequent targeted update which is fully described in \cite{rytgaard_one-step_2021}. 

Currently \pkg{concrete} can be used to estimate estimands derived from cause-specific absolute risks (e.g., risk ratios and risk differences) under static and dynamic interventions on binary treatments given at baseline. Estimands can be jointly targeted at multiple times, up to full risk curves over an interval, and for multiple events in cases with competing risks. Methods are available to handle right censoring, competing risks, and confounding by baseline covariates. Point estimates can be computed using g-formula plug-in or one-step TMLE, and asymptotic inference for the latter is derived from the variance of the efficient influence curve. 

\pkg{concrete} is not intended to be used for data with clustering, left trunctation (i.e. delayed entry) or interval censoring. Currently the Super Learners for estimating conditional hazards must be comprised of Cox regressions \citep{cox_regression_1972} although the incorporation of hazard estimators based on penalized Cox regressions and highly adaptive lasso are planned in future package versions. Support for stochastic interventions and interventions on multinomial and continuous treatments are also forthcoming, while longitudinal methods to handle time-dependent treatment regimes and time-dependent confounding are in longer term development.

\subsection{Other packages}
\label{otherpkgs}
\pkg{concrete} (\textbf{con}tinuous-time, \text{c}ompeting \textbf{r}isks, on\textbf{e}-step \textbf{t}argeted maximum likelihood \textbf{e}stimation) is the first R package on CRAN to implement a continuous-time TMLE for survival and competing risk estimands, but is related to existing R packages implementing semi-parametric efficient estimators for time-to-event-outcomes. The \CRANpkg{ltmle} \citep{schwab_ltmle_2020}, \href{https://github.com/osofr/stremr}{\pkg{stremr}} \citep{sofrygin_stremr_2017}, and \href{https://github.com/benkeser/survtmle}{\pkg{survtmle}} \citep{benkeser_survtmle_2019} implement discrete-time TMLEs for survival estimands and can be used to estimate right censored survival or competing risks estimands. All three packages implement an iterated expectations-based TMLE while \pkg{survtmle} also implements a discrete-time hazard-based TMLE formulation as well. \pkg{ltmle} and \pkg{stremr} can be applied to data structures with longitudinal treatment regimes and longitudinal confounding, which is an area of future development for \pkg{concrete}. 

Notably these packages all operate on a discrete time scale and would thus require discretization of time-to-event data observed continuously or near-continuously. While discretization of data with longitudinal confounding has been shown to negatively impact estimation \citep{sofrygin_targeted_2019, ferreira_guerra_impact_2020}, discretization choices (i.e. cutpoint location and number of intervals) have been shown to affect the performance of various discrete-time survival estimators even in the absence of longitudinal confounding \citep{argyropoulos_analysis_2015, sloma_empirical_2021, kvamme_continuous_2021, suresh_survival_2022, hickey_adaptive_2024}. Though discrete-time estimators on optimally discretized data generally performed at least as well as their continuous-time counterparts, the risk of degraded estimator performance has led authors to recommend treating discretization as a tuning parameter. This, to the best of our knowledge, is not standard practice in biomedical survival analyses and the aforementioned discrete-time TMLE packages do not provide built-in methods for optimizing discretization. Thus the value of concrete in the context of these pre-existing discrete-time TMLE packages is that it offers researchers with plausibly continuous survival data the chance to avoid this unnecessary potential complication.

In addition, the \ctv{CausalInference} CRAN Task View lists \CRANpkg{riskRegression} \citep{gerds_riskregression_2022} as estimating treatment effect estimands in survival settings using the inverse propensity of treatment weighted (IPTW) and double-robust augmented IPTW (AIPTW) estimators. None of the packages listed on the \ctv{Survival} CRAN Task View are described as implementing efficient semi-parametric estimators, though available via Github are the R packages \href{https://github.com/RobinDenz1/adjustedCurves}{\pkg{adjustedCurves}} \citep{denz_comparison_2023} and \href{https://github.com/tedwestling/CFsurvival}{\pkg{CFsurvival}} \citep{westling_inference_2021}, which implement the AIPTW and a cross-fitted doubly-robust estimator respectively. 

\subsection{Structure of this manuscript}
\label{structure}
This article is written for readers wishing to use the \pkg{concrete} package for their own analyses and for readers interested in an applied introduction to the one-step continuous-time TMLE method described in \citep{rytgaard_one-step_2021}. Section \ref{targetedlearning} outlines the targeted learning approach to time-to-event causal effect estimation, with subsection \ref{estimation} providing details on our one-step TMLE implementation. Usage of the \pkg{concrete} package and its features is then provided in Section \ref{UsingConcrete}, continuing the above example of a simple competing risks analysis of the PBC dataset. 

\section{The Targeted Learning framework for survival analysis}
\label{targetedlearning}
At a high level, the targeted learning roadmap for analyzing continuous-time survival or competing risks consists of:
\begin{enumerate}
  \item Specifying the causal model and defining a causal estimand (e.g. causal risk difference). Considerations include defining a time zero and time horizon, specifying the intervention (i.e., treatment) variable and the desired interventions (including on sources of right censoring), and specifying the target time(s) and event(s) of interest.
  \item Defining a statistical model and statistical estimand, and evaluating the assumptions necessary for the statistical estimand to identify the causal estimand. Considerations include identifying confounding variables, establishing positivity for desired interventions, and formalizing knowledge about the statistical model (e.g. dependency structures or functional structures).
  \item Performing estimation and providing inference. Considerations include prespecification of an estimator and an inferential approach with desirable theoretical properties (e.g. consistency and efficiency within a desired class), and assessing via outcome-blind simulations the estimator's robustness and suitability for the data at hand.
\end{enumerate}

In the following sections we discuss these three stages in greater detail.

\subsection{The causal model: counterfactuals, interventions, and causal estimands}
\label{CausalData}
With time-to-event data, typical counterfactual outcomes are how long it would take for some event(s) to occur if subjects were hypothetically to receive some intervention, i.e. treatment. Let \(A\) be the treatment variable and let \(d\) be the hypothetical intervention rule of interest, i.e., the function that assigns treatment levels to each subject. The simplest interventions are static rules setting \(A\) to some value \(a\) in the space of treatment values \(\mathcal{A}\), while more flexible dynamic treatment rules might assign treatments based on subjects' baseline covariates (which we denote as \(W\)), and stochastic treatment rules incorporate randomness and may even depend on the natural treatment assignment mechanism in so-called modified treatment policies. Additionally, our goal in time-to-event analyses is often to assess the causal effect of some treatment rule \(d\) on an event (or set of competing events) \textit{in the absence of right censoring}. This "absence of right censoring" condition is in fact a static intervention to deterministically prevent right censoring, and is an implicit component to many interventions in time-to-event analyses. 

Regardless of the type of intervention rule, the associated counterfactual survival data under intervention rule \(d\), \(X \sim P^d\), takes the general form
\begin{equation}
 X = \left(T^d,\, \Delta^d,\, A^d, \L \right) \label{causaldata}
\end{equation}
where \(T^d \in (0, t_{max}]\) is the counterfactual time-to-event under intervention \(d\) for the earliest of \(J\) competing events up to some maximum follow-up time \(t_{max}\), \(\Delta^d \in \{1, \dots, J\}\) is the counterfactual event index indicating which the \(J\) events would have hypothetically occurred first, and \(A^d\) is the treatment variable under intervention \(d\) (which for static and dynamic interventions will be a degenerate variable). Note that we differentiate between competing events (indexed \(1, ..., J\)) and sources of right censoring (not present in \(X\)), as our goal is to assess the causal effect of treatment rule \(d\) on the set of competing events in the absence of right censoring.  For ideal experiments tracking just one event, i.e. \(J = 1\), the causal setting is one of survival of a single risk; if instead mutually exclusive events would be allowed to compete, then the causal setting is one with competing risks. 

With the counterfactual data defined, causal estimands can then be specified as functions of the counterfactual data. For instance, if we were interested in effects of interventions \(d_0\) versus \(d_1\) on time-to-event outcomes, the counterfactual data \(\tilde{X} \sim P^{\,0,1}\) might be represented as
\begin{align*}
\tilde{X} = \left(T^{d_0},\, \Delta^{d_0},\, A^{d_0}, T^{d_1},\, \Delta^{d_1},\, A^{d_1}, \L \right)
\end{align*}
We could then define estimands such as the causal event \(j\) relative risks at time \(t\) 
\begin{align}
\tilde\Psi_{j, t}(P^{\,0,1}) = \frac{{P}(T^{d_1} \leq t, \Delta^{d_1} = j)}{{P}(T^{d_{0}} \leq t, \Delta^{d_{0}} = j)}
\label{causalrisk}
\end{align}
These estimands may be of interest at a single timepoint, at multiple timepoints, or over a time interval, and in the case of competing risks may involve multiple events (e.g. \(\tilde\Psi_{j, t}(P^{\,0, 1}) : t \in (0, t_{max}), \, j \in 1, \dots, J\)). In any case, once the desired causal quantity of interest has been expressed as a function of the counterfactual data, efforts can then be made to identify the causal estimand with a function of observed data, i.e. a statistical estimand.

\subsection{Observed data, identification, and statistical estimands}
\label{ObservedData}
Observed time-to-event data \(O \sim P_0\) with \(J\) competing events can be represented as:
\begin{equation}
 O = \left(\T,\, \tDelta,\, A,\, \L \right) \label{obs-data}
\end{equation}
where \(\T \in (0, t_{max}]\) is the earlier of the first event time \(T\) or the right censoring time \(C\), \(\tDelta \in \{0, \dots, J\}\) indicates which event occurs (with 0 indicating right censoring), \(A\) is the observed treatment and \(\L\) is the set of baseline covariates.

To link causal estimands such as Eq. \eqref{causalrisk} to statistical estimands, we need a set of identification assumptions to hold, informally: consistency, positivity, and conditional exchangeability (or their structural causal model analogs) . Readers can find a full discussion of these identification assumptions for absolute risk estimands in Section 3 of \citep{rytgaard_targeted_2022}. Given these assumptions, we can identify the cause-\(j\) absolute risk at time \(t\) under intervention \(d\) using the g-computation formula \citep{robins_new_1986} as
\begin{equation}
P(T^d \leq t, \Delta^d = j) = \mathbb{E}_{\mathcal{\L}} \left[ \int_{\mathcal{A}} \,  F_j(t \aX) \, \g^* (a \mid \L) \, da \right] \label{absrisk}
\end{equation}
where \(\g^*(a \mid \l)\) is the treatment propensity implied by the intervention \(d\). Here \(F_j(t \ax)\) is the conditional cause-\(j\) absolute risk
\begin{equation*}
F_j(t \ax) = \int_0^t \lambda_j(s \ax) \, S(s\texttt{-} \ax) \, ds \;,
\end{equation*}
where the cause-\(j\) conditional hazard function \(\lambda_j\) is defined as
\begin{equation*}
\lambda_j(t \ax) = \lim\limits_{h \to 0} \frac{1}{h} P(\T \leq t + h,\, \tDelta = j \mid \T \geq t,\, a,\, \x) \;,
\end{equation*}
and the conditional event-free survival probability is given by
\begin{equation}
S(t \ax) = \exp\left(-\int^{t}_{0} \sum\limits_{j=1}^{J} \lambda_j(s \ax) \, ds \right).\label{evfreesurv}
\end{equation}
From Eq \eqref{absrisk}, it follows that we can identify the causal cause-\(j\) relative risk \eqref{causalrisk} at time \(t\) by
\begin{equation}
\Psi_{F_{j,t}}(P_0) = \frac{\mathbb{E}_{\mathcal{\L}} \left[ \int_{\mathcal{A}} \,  F_j(t \aX) \, \g^*_{d_1} (a \mid \L) \, da \right]}{\mathbb{E}_{\mathcal{\L}} \left[ \int_{\mathcal{A}} \,  F_j(t \aX) \, \g^*_{d_0} (a \mid \L) \, da \right]} \label{obsrisk}
\end{equation}
where \(\g^*_{d_0}\) and \(\g^*_{d_1}\) represent the treatment propensities implied by treatment rules \(d_0\) and \(d_1\) respectively.

It should be noted that even without the identification assumptions for causal inference, statistical estimands such as Eq. \eqref{obsrisk} may still have valuable interpretations as standardized measures isolating the importance of the "intervention" variable \citep{laan_statistical_2006}.

\subsection{Targeted estimation}
\label{estimation}
The TMLE procedure for estimands derived from cause-specific absolute risks begins with estimating the treatment propensity \(\g\), the conditional hazard of censoring \(\lambda_c\) and the conditional hazards of events \(\lambda_j \,:\; j = 1, \dots, J\). In \pkg{concrete} these nuisance parameters are estimated using the Super Learner algorithm, which involves specifying a cross-validation scheme, compiling a library of candidate algorithms, and designating a cross-validation loss function and a Super Learner meta-learner.

\subsubsection{Estimating treatment propensity}
\label{trtps-est}
Let \(\g_0\) be the true conditional distribution of \(A\) given \(\X\) (i.e. the treatment propensity), let
\(\mathcal{M}_{\g} = \left\{\Hat{\g} : P_n \to \Hat{\g}(P_n)\right\}\)
be the library of candidate propensity score estimators, and let \(L_\g\) be a loss function such that the risk \(\mathbb{P}_0\,L_\g(\g) \equiv \mathbb{E}_0\left[L_\g(\g, O)\right]\) is minimized by \(\g_0\). The discrete Super Learner estimator is then the candidate propensity estimator with minimal cross validated risk, 
\begin{equation}
\Hat{\g}^{SL} = \argmin_{\Hat{\g} \in \mathcal{M}_\g} \sum_{v = 1}^{V} \mathbb{P}_{Q^\mathcal{V}_v} \; L_\g(\Hat{\g}(P^\mathcal{T}_v)) \label{propsl}
\end{equation}
where \(\Hat{\g}(P^\mathcal{T}_v)\) are candidate propensity score estimators trained on data \(Q^\mathcal{T}_v\). Currently \pkg{concrete} uses default \CRANpkg{SuperLearner} \citep{polley_superlearner_2021} loss functions (non-negative least squares) and with a default Super Learner library consisting of elastic-net and extreme gradient boosting. 

\subsubsection{Estimating conditional hazards}
\label{haz-est}
For \(\delta = 0, \dots, J\) where (\(\delta = 0\)) is censoring and (\(\delta \in \{1, \dots, J\}\)) are outcomes of interest, let \(\lambda_{\delta} \,:\; \delta = 0, \dots, J\) be the true conditional hazards, let \(\mathcal{M}_\delta = \{\Hat{\lambda}_\delta : P_n \to \Hat\lambda_{\delta}(P_n)\}\) be the libraries of candidate estimators, and let \(L_{\delta}\) be loss functions such that the risks \(\mathbb{P}_0\,L_{\delta}(\cdot)\) are minimized by the true conditional hazards \(\lambda_{\delta}\). The discrete Super Learner selectors for each \(\delta\) then chooses the candidate which has minimal cross validated risk 
\begin{equation}
\Hat{\lambda}_\delta^{SL} = \argmin_{\Hat{\lambda}_\delta \in \mathcal{M}_\delta} \sum_{v = 1}^{V} \mathbb{P}_{Q^\mathcal{V}_v} \; L_{ \delta}(\Hat{\lambda}_{\delta}(P^\mathcal{T}_v)) \;:\; \delta = 0, \dots, J\label{hazsl}
\end{equation}
where \(\Hat{\lambda}_\delta(P^\mathcal{T}_v)\) are candidate event \(\delta\) conditional hazard estimators trained on data \(Q^\mathcal{T}_v\). The current \pkg{concrete} default is a library of two Cox models, treatment-only and main-terms, with cross-validated risk computed using negative log Cox partial-likelihood loss \citep{cox_partial_1975, rytgaard_targeted_2022}
\[ \mathbb{P}_{Q^\mathcal{V}_v} \; L_{ \delta}(\Hat{\lambda}_{\delta}(P^\mathcal{T}_v)) =  \mathbb{P}_{Q^\mathcal{V}_v} \; L_{ \delta}(\Hat{\beta}_{\delta, Q^\mathcal{T}_v}) = - \sum_{i: \, O_i \in Q^\mathcal{V}_v} \left[\Hat{\beta}^{'}_{\delta, Q^\mathcal{T}_v}\,\L_i - \log\left[\sum_{i: \, O_i \in Q^\mathcal{V}_v} \mathbb{1}\left(\T_l \geq \T_i\right) \exp(\Hat{\beta}^{'}_{\delta, Q^\mathcal{T}_v}\,\L_h)\right]\right] \,\]
where \(\L_{h}\) are the covariates of the risk set at time \(t\), \(\{h \,:\, \T_h \geq t\}\) and \(\Hat{\beta}^{}_{\delta, Q^\mathcal{T}_v}\) are the coefficients of an event \(\delta\) candidate Cox regression trained on data \(Q^\mathcal{T}_v\). 

\subsubsection{Solving the efficient influence curve equation}
\label{EIC}
For parameters such as risk ratios which are derived from cause-specific absolute risks, we solve a vector of absolute risk efficient influence curve (EIC) equations with one element for each combination of target event, target time, and intervention. That is, the EIC for a target parameter involving \(J\) competing events, \(K\) target times, and \(M\) interventions is a \(J \times K\times M\) dimensional vector where the component corresponding to the cause-specific risk of event \(\jj\), at time \(t_k\), and under intervention propensity \(\trt_{m}\) is:
\begin{align}
    D^*_{m, \jj, k}(\lambda, \g, S_c)(O) = \sum_{\lj = 1}^{J} \int \; &h_{m,\, \jj,\, k,\, \lj,\, s}(\lambda, \g, S_c)(O) \, \left(N_{\lj}(s) - \1(\T \geq s) \, \lambda_\lj(s \AX)\right) \, ds \label{eic} \\
    &{
    + \int_{\mathcal{A}} F_\jj(t_k \mid A = a, \X)\,\trt_m(a \mid \X) \, da - \Psi_{\trt, \jj, t}(P_0)}  \nonumber 
\end{align}
where \(N_l : l = 0,\dots, J\) are the cause-specific counting processes
\[N_l(s) = \1\left\{\T \leq s, \tDelta = l\right\} \]
and \(h_{m,\, \jj,\, k,\, \lj,\, s}(\lambda, \g, S_c)(O)\) is the TMLE "clever covariate"
\begin{align}
    h_{m,\, \jj,\, k,\, \lj,\, s}&(\lambda, \g, S_c)(O) = \frac{{\color{blue}\trt_m(A \mid \X)\,} \1(s \leq t_k)}{{\color{green!70!black}\g(A \mid \X) \;S_c(s\text{-} \AX)}} \, \bigg(\1(\lj = \jj) - \frac{{\color{red}F_\jj(t_k \AX)} - {\color{red} F_\jj(s \AX)}}{{\color{red} S(s \AX)}}\bigg) \label{clevcov}
\end{align}
We highlight here that the clever covariate is a function of the {\color{blue}intervention-defined treatment propensity}, the {\color{green!70!black}observed intervention-related densities} (i.e. the observed treatment propensity and cumulative conditional probability of remaining uncensored) which are unaffected by TMLE targeting, and the {\color{red}observed outcome-related densities} which will be updated by TMLE targeting. Note also that our notation for the EIC (\(D^*_{m, \jj, k}(\lambda, \g, S_c)(O)\)) and clever covariate (\(h_{m,\, \jj,\, k,\, \lj,\, s}(\lambda, \g, S_c)(O)\)) reflects the dependence on \(P\) through the cause-\(j\) conditional hazards \(\lambda = (\lambda_l \;:\;  l = 1, \dots, J)\) and the treatment propensity \(\g\) and conditional censoring survival \(S_c(t \ax) = \exp\left(-\int^{t}_{0} \lambda_0(s \ax) \, ds \right)\). 

The one-step continuous-time survival TMLE \citep{rytgaard_one-step_2021} involves updating the cause-specific hazards \(\lambda\) along the universally least favorable submodel, which is implemented as small recursive updates along a sequence of locally least favorable submodels. To describe this procedure, let us first introduce the following vectorized notation:
\begin{align*}
{D}^{*} &= \left(D^*_{m, \jj, k} : m = 1,\dots,M \,,\; \jj=1,\dots,J \,,\; k=1,\dots,K\right)\\
h_{\lj, s} &= \left(h_{m,\, \jj,\, k,\, \lj,\, s} : m = 1,\dots,M \,,\; \jj=1,\dots,J \,,\; k=1,\dots,K\right)
\end{align*}
The one-step continuous-time survival TMLE recursively updates the cause-specific hazards in the following manner: starting from \(b=0\), with \(\lambda^0_j = \hat{\lambda}^{SL}_j\), and \(\lambda^b = \left(\lambda^b_l \;:\; l = 1, \dots, J\right)\)
\begin{equation}
\lambda^{b+1}_{l} = \lambda^{b}_l \, \exp \left( \frac{\left<\mathbb{P}_n {D}^*( \lambda^b, \g,  S_c)(O),\; h_{j, s}( \lambda^b, \g,  S_c)(O) \right>}{|| \mathbb{P}_n {D}^*( \lambda^b, \g, S_c)(O)||} \; \epsilon_b\right), \quad l = 1,\dots,J \label{one-step}
\end{equation}

where
\begin{align*}
\left<x , y \right>& = x^\top y \hspace{.5cm}, \hspace{.5cm} ||x|| = \sqrt{x^\top x}
\end{align*}
and the step sizes \(\epsilon_b\) are chosen such that
\[|| \mathbb{P}_n {D}^*( \lambda^{b+1}, \g, S_c)(O)|| < || \mathbb{P}_n {D}^*( \lambda^{b}, \g, S_c)(O)|| \;\;. \]
The recursive update following Eq \eqref{one-step} is completed at the iteration \(B\) where
\begin{equation}
\left|\mathbb{P}_n {D}^*( \lambda^B, \g, S_c)(O)\right| \leq \frac{\sqrt{\mathbb{P}_n {D}^*( \lambda^B, \g, S_c)(O)^2}}{\sqrt{n} \, \log(n)} \label{one-step-stop}
\end{equation}
This updated vector of conditional hazards \(\lambda^B\) is then used to compute a plug-in estimate of the statistical estimand simultaneously across causes, target times, and interventions. 

\subsubsection{Estimating variance}
\label{variance}

In \pkg{concrete}, the variance of the TMLE is estimated based on the plug-in estimate of the sample variance of the EIC, \(\frac{\mathbb{P}_n \;D^*( \hat\lambda^B, \hat\g, \hat S_c)(O)^2}{n}\), which is a consistent estimator for the variance of the TMLE when all nuisance parameter estimators are consistent. In the presence of practical positivity violations arising from sparsity in finite samples (discussed further in Section \ref{concreteest}), the EIC-based variance estimator can be anti-conservative and variance estimation by bootstrap may be more reliable \citep{tran_robust_2018}. However, bias resulting from positivity violations cannot be remedied in this way, and so other methods of addressing positivity violations are recommended instead \citep{petersen_diagnosing_2012}. For multidimensional estimands, simultaneous confidence intervals can be computed by simulating the \(1 - \alpha\) quantiles of a multivariate normal distribution with the covariance structure of the estimand EICs.

\subsubsection{Specifying a Super Learner}
\label{cv}
For a simple \(V\text{-fold}\) cross-validation setup, let 
\(Q_n = \{O_i\}_{i=1}^n \sim P_n\) 
be the observed \(n\) i.i.d observations of \(O \sim P_0\) and let
\(B_n \in \{1, ... , V\}^n\)
be a random vector that assigns the \(n\) observations into \(V\) validation folds. Then for each \(v\) in \(1, ..., V\) we define a training set 
\(Q^\mathcal{T}_v = \{O_i : B_n^i = v\} \sim  P^\mathcal{T}_v\)
and corresponding validation set
\(Q^\mathcal{V}_v = \{O_i : B_n^i \neq v\} \sim P^\mathcal{V}_v\).

Having specified a cross-validation scheme, the next steps are to construct the Super Learner candidate library, define an appropriate loss function, and select a Super Learner meta-learner. Super Learner libraries should be comprised of candidate algorithms that range in flexibility while respecting existing data-generating knowledge. For instance, candidate estimators should incorporate domain knowledge regarding covariates and interactions that are predictive of outcomes. If the number independent observations \(n\) is small compared to the number of covariates, then Super Learner libraries should contain fewer candidates and either incorporate native penalization, e.g. regularized Cox regression (\code{coxnet}) \citep{simon_regularization_2011}, or be paired with covariate screening algorithms. If, on the other hand, the number of independent observations is large compared to the number of covariates, then Super Learner libraries can include more algorithms including highly flexible non-parametric algorithms such as highly adaptive lasso (HAL). It should be noted that using HAL for initial nuisance parameter estimation can achieve the necessary convergence rates \citep{laan_generally_2017, bibaut_fast_2019, rytgaard_estimation_2023} for TMLE to be efficient. Super Learner loss functions should imply a risk that is minimized by the true data-generating process and define a loss-based dissimilarity tailored to the target parameter and a discrete selector that selects the best performing candidate should be used as the Super Learner meta-learner \citep{laan_super_2007}. If computationally feasible, Super Learners using more flexible meta-learner algorithms can safely be nested as candidates within a larger Super Learner using a discrete meta-learner. Additional guidance on Super Learner specification is provided in \citep{phillips_practical_2023} and Chapter 3 of \citep{laan_targeted_2011}.

Currently the default cross-validation setup in \pkg{concrete} follows the guidelines laid out in \citep{phillips_practical_2023}, with the number of cross-validation folds increasing with decreased sample size. The default number of folds ranges from leave-one-out cross validation (LOOCV) for datasets with fewer than 30 independent observations to 2-fold cross validation for datasets with over 10000 independent observations. Default Super Learner libraries are provided and will be detailed in the following sections, but should be amended to suit the data at hand and to incorporate subject matter knowledge.

\section{Using concrete}
\label{UsingConcrete}
The basic \pkg{concrete} workflow consists of using three functions sequentially: 
\begin{itemize}
\item \code{formatArguments()}
\item \code{doConcrete()}
\item \code{getOutput()}.
\end{itemize}
Users specify their estimation problem and desired analysis through \code{formatArguments()}, which checks the specified analysis for potential issues and produces a \code{"ConcreteArgs"} object containing the specification of the target estimand and the continuous-time one-step survival TMLE. The \code{"ConcreteArgs"} object is then passed into \code{doConcrete()} which performs the specified estimation and produces a \code{"ConcreteEst"} object which can be interrogated for diagnostics and intermediate estimation outputs such as initial nuisance parameter estimates. The \code{"ConcreteEst"} object can then be passed into \code{getOutput()} to produce tables and plots of cause-specific absolute risk derived estimands such as risk differences and relative risks. 

\subsection{Defining the estimation problem and specifying the estimator}
\label{formatArguments}

Broadly speaking, the arguments of \code{formatArguments()} are involved in specifying the data structure, the target estimand, and the TMLE estimator. The output of \code{formatArguments()}, a \code{"ConcreteArgs"} object, contains all of the necessary details to specify a continuous-time TMLE analysis, can be printed provide a summary of the specified estimation targets and estimator, and can be iteratively modified as the user refines their target estimand and estimator.

\begin{verbatim}
ConcreteArgs <- formatArguments(
    # Data #
    DataTable,		# data.frame or data.table
    EventTime,      	# name of event time variable
    EventType,     	# name of event status variable
    Treatment,        	# name of treatment variable
    # Estimand #
    Intervention,      	# 2 static interventions
    TargetTime, 		# 7 target times: 3-6 years biannually
    TargetEvent,     	# 2 competing risks
    # Estimator #
    CVArg,         		# 10-Fold Cross-Validation
    Model,			# using default Super Learner libraries
)
\end{verbatim}

\subsubsection{Data}
\label{ObservedDataConcrete}
The observed data are passed into the \code{DataTable} argument as either a \code{data.frame} or \code{data.table} object, which must contain columns corresponding to the observed time-to-event \(\T\), the indicator of which event occured \(\Delta\), and the treatment variable \(A\). Treatment values in \(A\) must be numeric, with binary treatments encoded as 0 and 1, and if the dataset contains an ID column, its name should be passed into the \code{ID} argument. Any number of columns containing baseline covariates \(\L\) can also be included. All data inputs must be without missingness; imputation of missing covariates should be done prior to passing data into \pkg{concrete} while missing treatment or outcome values, aside from right-censoring, is not supported. 

By default \pkg{concrete} pre-processes covariates uses \code{model.matrix()} to one-hot encode factor variables in order to facilitate compatibility between candidate regression implementations which may process categorical variables differently. The \code{"ConcreteArgs"} object returned by \code{formatArguments()} includes the reformatted data as \code{.[["DataTable"]]} and the mapping of new covariate names to the originals can be retrieved by calling \code{attr(.[["DataTable"]], "CovNames"}.This pre-processing can be turned off by setting \code{RenameCovs = FALSE}, which can be important for specifying dynamic interventions as will be discussed in the next section.

\subsubsection{Target estimand: intervention, target events, and target times}
\label{Estimand}
Static interventions on a binary treatment \(A\), i.e. setting all observations to \(A=0\) or \(A=1\), can specified by setting \code{Intervention} to 0 or  1 respectively. If both interventions are of interest, i.e. for contrasts such as risk ratios and risk differences, then \code{Intervention} should be set to c(0, 1). More complex interventions can be specified with a list containing a pair of functions: an "intervention" function which outputs desired treatment assignments and a "g.star" function which outputs desired treatment probabilities. "intervention" functions take three data.table inputs: the first containing treatment column(s), the second containing baseline covariates, and the third containing the propensity scores for observed treatment values. The "intervention" function output must be a data.table containing the desired intervention values, with the same dimensions and column names as the input treatment data.table. "g.star" functions take an additional fourth data.table argument containing intervention values (i.e. the output of the "intervention" function) and must return a data.table containing the intervention treatment probabilities which has the same dimensions and column names as the intervention values data.table. The function \code{makeITT()} creates list of functions corresponding to the binary treat-all and treat-none static interventions, which can be used as a template for specifying more complex interventions. When specifying dynamic interventions using covariate names, it may be important to set \code{RenameCovs = FALSE}, as otherwise \pkg{concrete} would potentially rename factor covariates. . 

The \code{TargetEvent} argument specifies the event types of interest. Event types must be be coded as integers, with non-negative integers reserved for censoring. If \code{TargetEvent} is left \code{NULL}, then all positive integer event types in the observed data will be jointly targeted. In the \code{pbc} dataset, there are 3 event values encoded by the\code{status} column: 0 for censored, 1 for transplant, and 2 for death. To analyze \code{pbc} with transplants treated as right censoring, \code{TargetEvent} should be set to 2, whereas for a competing risks analysis one could either leave \code{TargetEvent = NULL} or set \code{TargetEvent = 1:2} as in the above example.

The \code{TargetTime} argument specifies the times at which the cause-specific absolute risks or event-free survival are estimated. Target times should be restricted to the time range in which target events are observed and \code{formatArguments()} will return an error if target time is after the last observed failure event time. If no \code{TargetTime} is provided, then \pkg{concrete} will target the last observed event time, though this is likely to result in a highly variable estimate if prior censoring is substantial. The \code{TargetTime} argument can either be a single number or a vector, as one-step TMLE can target cause-specific risks at multiple times simultaneously. 

\subsubsection{Estimator specification} \label{EstimationSpec}
The arguments of \code{formatArguments()} involved in estimation are the cross-validation setup \code{CVArg}, the Super Learner candidate libraries \code{Model}, the software backends \code{PropScoreBackend} and \code{HazEstBackend}, and the practical TMLE implementation choices \code{MaxUpdateIter}, \code{OneStepEps}, and \code{MinNuisance}. Note that \code{Model} is used in this section in line with common usage in statistical software, rather than to refer to formal statistical or causal models as in preceding sections. 

Cross-validation is implemented by calling \code{origami::make\_folds()} with the \code{CVArg} argument. If no input is provided into \code{CVArg}, the default cross-validation setup follows the recommendations in \citep{phillips_practical_2023}. Cross-validation folds are stratified by event type and the number of folds ranges from 2 for datasets with greater than 10000 independent observations to LOOCV for datasets with fewer than 30 independent observations. Chapter 5 of the online Targeted Learning Handbook \citep{malenica_chapter_nodate} demonstrates the specification of several other cross-validation schemes.

Super Learner libraries for estimating nuisance parameters are specified through the \code{Model} argument. The input should be a named list with an element for the treatment variable and one for each event type including censoring as illustrated in the following code example. The list element corresponding to treatment must be named as treatment variable, and the list elements corresponding to each event type must be named with the corresponding event type value (e.g. "0" for censoring). Any missing specifications will be filled in with defaults, and the resulting list of libraries can be accessed in the output \code{.[["Model"]]} and further edited by the user, as shown below.

\begin{verbatim}
# specify regression models
ConcreteArgs$Model <- list(
    "trt" = c("SL.glmnet", "SL.bayesglm", "SL.xgboost", "SL.glm", "SL.ranger"),
    "0" = NULL, # will use the default library
    "1" = list(Surv(time, status == 1) ~ trt, Surv(time, status == 1) ~ .),
    "2" = list("Surv(time, status == 2) ~ trt", "Surv(time, status == 2) ~ .")
)
\end{verbatim}

In \pkg{concrete}, propensity scores are by default estimated using the candidate algorithms \code{c("xgboost", "glmnet")} implemented by packages \CRANpkg{xgboost} \citep{chen_xgboost_2022} and \CRANpkg{glmnet} \citep{friedman_regularization_2010}. For further details about these packages, see their respective package documentations.

For estimating the necessary conditional hazards, \pkg{concrete} currently relies on a discrete Superlearner consisting of a library of Cox models implemented by \code{survival::coxph()} evaluated on cross-validated partial-likelihood loss as detailed in Section \ref{haz-est}. Support for estimation of hazards using coxnet \citep{simon_regularization_2011}, Poisson-HAL and other methods is planned in future package versions. The default Cox specifications are a treatment-only regression and a main-terms regression including treatment and all covariates. These models can be specified as strings or formulas as can be seen in the above example.

As detailed by Eq. \eqref{one-step} and \eqref{one-step-stop}, the one-step TMLE update step involves recursively updating cause-specific hazards, summing along small steps, scaled by a multiplicative factor \(\epsilon_b\). The default initial scaling factor is 0.1, and each time an update step would not decrease the mean estimated EIC, the step size scaling factor is halved and the update step is re-tried. The \code{MaxUpdateIter} argument is used to provide a definite stop to the recursive TMLE update. The default of 500 steps should be sufficient for most applications, but may need to be increased when targeting estimands with many components or for rare events. The presence of practical positivity sparsity can also result in slow TMLE convergence, but increasing \code{MaxUpdateIter} would not be not an adequate solution there as the resulting TMLE estimates and inference may still be unreliable.
The \code{MinNuisance} argument specifies a lower bound, with a mirrored \code{1 - MinNuisance} upper bound, for the product of the propensity score and lagged survival probability for remaining uncensored; this term is present in the denominator of the efficient influence function and bounding improves estimator stability at the cost of introducing bias.

\subsubsection{Modifying the specified estimation}
\label{concreteargs}
The \code{"ConcreteArgs"} output of \code{formatArguments()} is an environment containing the estimation specification which can then be modified by the user. Modified \code{"ConcreteArgs"} object should then be passed back through \code{formatArguments()} to check the updated estimation specification. 

\begin{verbatim}
# decrease the maximum tmle update number to 50
ConcreteArgs$MaxUpdateIter <- 50

# add a candidate regression with treatment interactions
ConcreteArgs[["Model"]][["2"]][[3]] <- "Surv(time, status == 2) ~ trt*."

# validate new estimation specification
ConcreteArgs <- formatArguments(ConcreteArgs)
\end{verbatim}

\code{"ConcreteArgs"} objects can be printed to display summary information about the specified estimation problem,

\begin{verbatim}
print(ConcreteArgs, Verbose = FALSE)
\end{verbatim}

Below, we can see that the specified analysis is for two competing risks (Target Events: 1, 2) under interventions "A=1" and "A=0" assigning all subjects to treated and control arms, respectively. Objects in the \code{"ConcreteArgs"} environment can be interrogated directly for details about any particular aspect of the estimation specification. For instance, as mentioned before the one-hot encoding of covariates can be seen in a table by \code{attr(.[["DataTable"]], "CovNames")}, intervention treatment assignments can be checked at \code{.[["Regime]]}.

\begin{figure}[H]
\includegraphics[width=\linewidth]{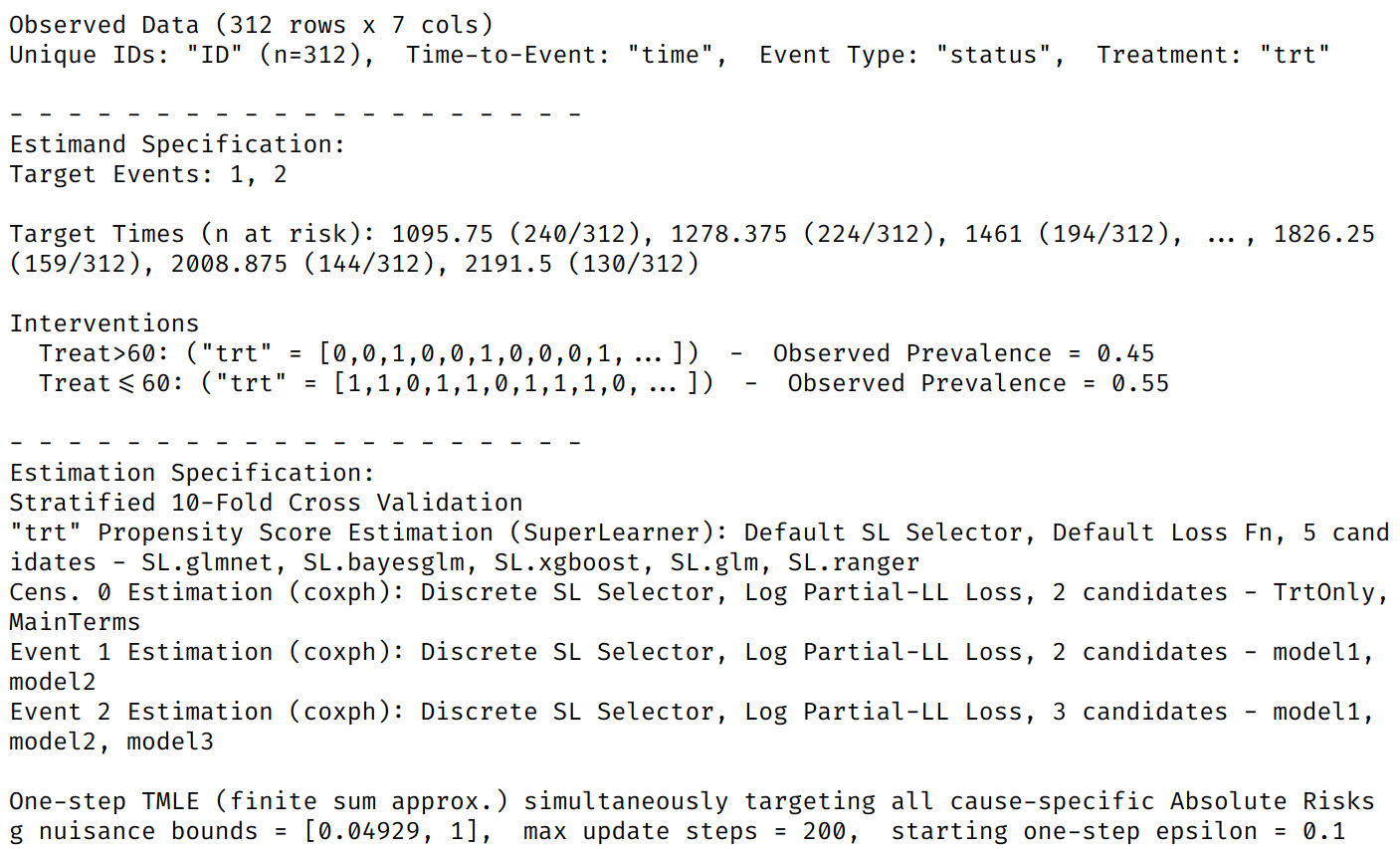}
\end{figure}

\subsection{Estimation}
\label{doConcrete}
To implement a specified analysis, \code{"ConcreteArgs"} objects are passed into the \code{doConcrete()} function which performs the specified TMLE analysis. The output is an object of class \code{"ConcreteEst"} which contains TMLE point estimates and corresponding estimated influence curves for the cause-specific absolute risks for each targeted event at each targeted time under each intervention. If the \code{GComp} argument is set to \code{TRUE}, then a Super Learner-based g-formula plugin estimate of the targeted risks will be included in the output. 

\begin{verbatim}
ConcreteEst <- doConcrete(ConcreteArgs)
\end{verbatim}

We have reviewed the one-step continuous-time TMLE implementation in Section \ref{estimation}, so here we will name the non-exported functions in \code{doConcrete()} which perform each of the steps of the one-step continuous-time survival TMLE procedure, in case users wish to explore the implementation in depth.

The initial estimation of nuisance parameters and is performed by the function \code{getInitialEstimate()} which depends on \code{getPropScore()} for propensity scores (Section \ref{trtps-est}) and \code{getHazEstimate()} for the conditional hazards (Section \ref{haz-est}).

Computing of EICs is done by \code{getEIC()} which is used within the \code{doTmleUpdate()} function which performs the one-step TMLE update procedure (Section \ref{EIC}).

\subsubsection{ConcreteEst objects}
\label{concreteest}
The print method for \code{"ConcreteEst"} objects summarizes the estimation target and displays diagnostic information about TMLE update convergence, intervention-related nuisance parameter bounding, and the nuisance parameter Super Learners.

\begin{verbatim}
    print(ConcreteEst, Verbose = FALSE)
\end{verbatim}
\begin{figure}[H]
\center
\includegraphics[width=\linewidth]{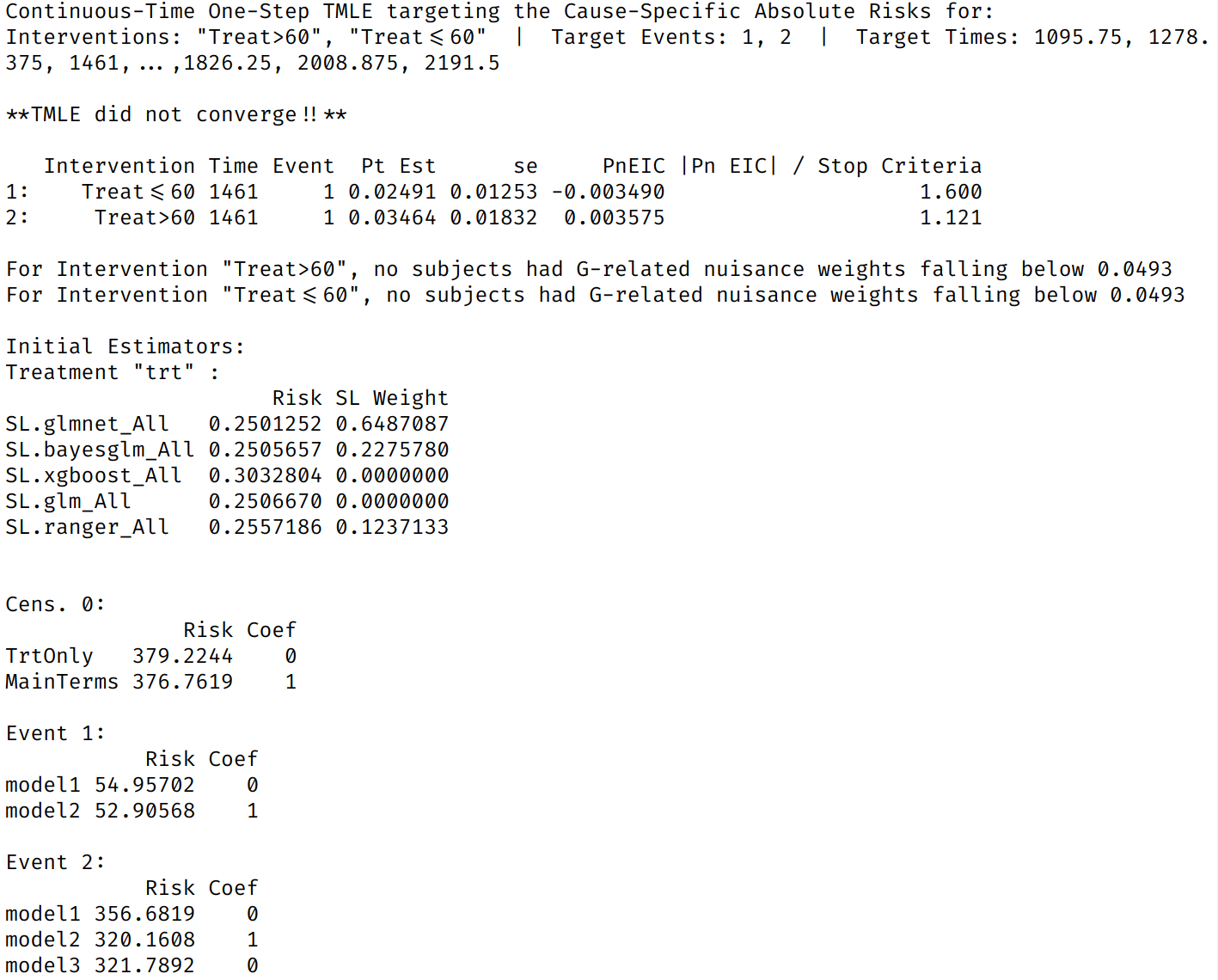}
\end{figure}

If TMLE has not converged, any mean EIC elements that have not attained the desired cutoff, i.e. Eq \eqref{one-step}, will be displayed in a table. For instance, we can see above that the absolute value of the mean EIC for intervention \samp{Treat>60} at time 1461 for event 1 has not reached the stopping criteria and is $1.6$ times larger than the stopping criteria. Increasing the the maximum number of TMLE update iterations via \code{MaxUpdateIter} can allow TMLE to finish updating nuisance parameters, though at target time points when few events have yet occurred even small mean EIC values may not meet the convergence criteria and adequate convergence may require many iterations. 

The extent to which the intervention-related nuisance parameters (i.e. propensity scores and probabilities of remaining uncensored) have been lower-bounded is also reported for each intervention both in terms of the percentage of nuisance weights that have been bounded and the percentage of subjects with bounded nuisance weights. If users suspect possible positivity issues, the plot method for \code{"ConcreteEst"} objects can be used to visualize the distribution of estimated propensity scores for each intervention, with the red vertical line marking the cutoff for lower-bounding.

\begin{verbatim}
    plot(ConcreteEst, ask  = FALSE)
\end{verbatim}

\begin{figure}[H]
\includegraphics[width=\linewidth]{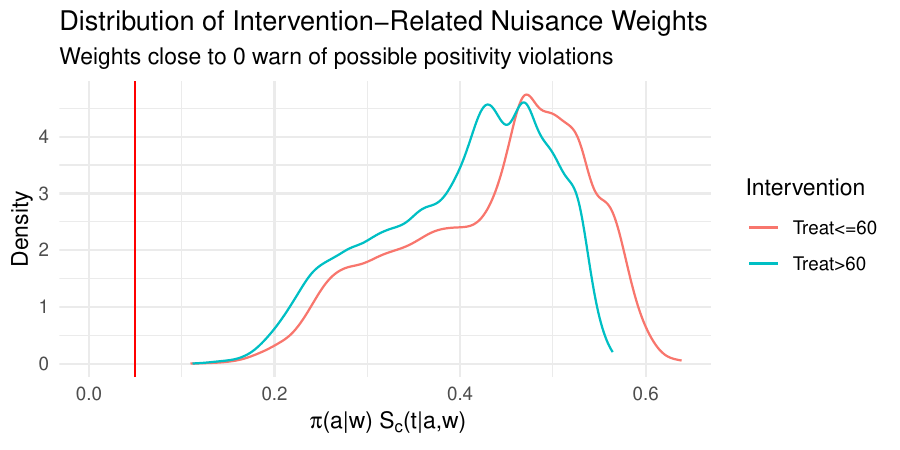}
\caption{}
\label{fig:ps-diag}
\end{figure}

Intervention-related nuisance parameters with values close to 0 indicate the possibility of positivity violations and may warrant re-examining the target time(s), interventions, and covariate adjustment sets. In typical survival applications, positivity issues may arise when targeting times at which some subjects are highly likely to have been censored, or if certain subjects are unlikely to have received a desired treatment intervention. As positivity violations not only impact causal interpretability, but also estimator behaviour, we urge users to re-consider their target analyses; \citep{petersen_diagnosing_2012} provides guidance on reacting to positivity issues.

The last three tables ("Cens. 0", "Event 1", and "Event 2") in the above code output show the candidate estimators of nuisance parameters, summarized with the cross-validated risk of each candidate estimator followed by their weight in the corresponding Super Learners. 

\subsection{Producing outputs}
\label{getoutput}
\code{getOutput()} takes as an argument the \code{"ConcreteEst"} object returned by \code{doConcrete()} and can be used to produce tables and plots of the cause-specific risks, risk differences, and relative risks. By default \code{getOutput()} returns a \code{data.table} with point estimates and pointwise standard errors for cause-specific absolute risks, risk differences, and risk ratios. By default, the first listed intervention is used as the "treated" group while the second is considered "control"; other contrasts can be specified via the \code{Intervention} argument. Below we show a subset of the relative risk estimates produced by the "nutshell" estimation specification for the pbc dataset. 

\begin{verbatim}
ConcreteOut <- getOutput(ConcreteEst = ConcreteEst, 
                         Estimand = "RD",
                         Intervention = 1:2, 
                         GComp = TRUE, 
                         Simultaneous = TRUE, 
                         Signif = 0.05)
head(ConcreteOut, 12)
\end{verbatim}

\begin{verbatim}
#>        Time Event  Estimand             Intervention Estimator   Pt Est    se
#> 1: 1095.750     1 Risk Diff [Treat>60] - [Treat<=60]      tmle  0.00800 0.018
#> 2: 1095.750     1 Risk Diff [Treat>60] - [Treat<=60]     gcomp  0.00300    NA
#> 3: 1095.750     2 Risk Diff [Treat>60] - [Treat<=60]      tmle -0.02000 0.040
#> 4: 1095.750     2 Risk Diff [Treat>60] - [Treat<=60]     gcomp  0.00025    NA
#> 5: 1278.375     1 Risk Diff [Treat>60] - [Treat<=60]      tmle  0.00800 0.018
#> 6: 1278.375     1 Risk Diff [Treat>60] - [Treat<=60]     gcomp  0.00300    NA
#>    CI Low CI Hi SimCI Low SimCI Hi
#> 1: -0.027 0.043    -0.039    0.055
#> 2:     NA    NA        NA       NA
#> 3: -0.099 0.058    -0.130    0.086
#> 4:     NA    NA        NA       NA
#> 5: -0.027 0.043    -0.039    0.055
#> 6:     NA    NA        NA       NA
\end{verbatim}

From left to right, the first five columns describe the estimands (target times, target events, estimands, and interventions) and estimators. The subsequent columns show the point estimates, estimated standard error, confidence intervals and simultaneous confidence bands. The \code{Signif} argument (set to a default of 0.05) specifies the desired double-sided alpha which is then used to compute confidence intervals, and the \code{Simultaneous} argument specifies whether or not to compute a simultaneous confidence band for all output TMLE estimates. Here we also see that when estimands involve many time points or multiple events, tables may be difficult to interpret at a glance. Instead plotting can make treatment effects and trends more visually interpretable, as was shown in Figure \ref{fig:nutshell-pbc-rd}.

The \code{plot} method for \code{"ConcreteOut"} object invisibly returns a list of \code{"ggplot"} objects, which can be useful for personalizing graphical outputs. Importantly, users should note that plots do not currently indicate if TMLE has converged or if positivity may be an issue; users must therefore take care to examine the diagnostic output of the \code{"ConcreteEst"} object prior to producing effect estimates using \code{getOutput()}.

\subsection{A \pkg{concrete} example: analyzing the competing risks in the PBC dataset}
\label{nutshell}
Below we illustrate the usage of \pkg{concrete} on the well-known Mayo Clinic Primary Biliary Cholangitis (PBC) data set \citep{fleming_counting_1991, therneau_modeling_2000}. We estimate the cause-specific counterfactual absolute risk differences, i.e. average treatment effects, under two levels of a binary treatment (randomization to placebo or D-penicillamine). The treatment column \code{"trt"} is transformed so that 0 indicates placebo and 1 indicates D-penicillamine, and where the two competing events are transplant (\code{"status"=1}) and death (\code{"status"=2}) in the presence of right censoring (\code{"status"=0}). We include outcomes for two estimators, g-computation plug-in and TMLE, as well as point-wise 95\% confidence intervals based on the estimated influence curves and 95\% simultaneous confidence bands for the treatment effects across all targeted time points. 

\subsubsection{Defining the problem}
\begin{verbatim}
ConcreteArgs <- formatArguments(
    DataTable = data,               # data.frame or data.table
    EventTime = "time",             # name of event time variable
    EventType = "status",           # name of event status variable
    Treatment = "trt",              # name of treatment variable
    ID = NULL,                      # (optional) name of the ID variable if present in input data 
    Intervention = 0:1,             # 2 static interventions
    TargetTime = 365.25/2 * (6:12), # 7 target times: 3-6 years biannually
    TargetEvent = 1:2,              # 2 competing risks
    CVArg = list(V = 10),           # 10-Fold Cross-Validation
    Model = NULL,                   # using default Super Learner libraries
    Verbose = FALSE                 # less verbose warnings and progress messages
)
\end{verbatim}

In the PBC example, the observed data is the \code{data} object, \(\T\) is the column \code{"time"}, \(\Delta\) is the column \code{"status"}, \(A\) is the column \code{"trt"}, and covariates \(L\) are the remaining columns: (\code{"age"}, \code{"sex"}, and \code{"albumin"}).

By default \pkg{concrete} pre-processes covariates using one-hot encoding to facilitate compatibility between candidate regression implementations which may process categorical variables differently. The \code{"ConcreteArgs"} object returned by \code{formatArguments()} includes the reformatted data as \code{.[["DataTable"]]} and the mapping of new covariate names to the originals can be retrieved by calling \code{attr(.[["DataTable"]], "CovNames"}.

\begin{verbatim}
attr(ConcreteArgs[["DataTable"]], "CovNames")
\end{verbatim}

\begin{verbatim}
#>    ColName CovName CovVal
#> 1:      L1     age      .
#> 2:      L2 albumin      .
#> 3:      L3     sex      f
\end{verbatim}

This pre-processing can be turned off by setting \code{RenameCovs = FALSE}, which can be important for specifying dynamic interventions as will be discussed in the next section.

\subsubsection{Target estimand: intervention, target events, and target times}
\label{Estimand}
Static interventions on a binary treatment \(A\) setting all observations to \(A=0\) or \(A=1\) can specified with 0, 1, or c(0, 1) if both interventions are of interest, i.e. for contrasts such as risk ratios and risk differences. More complex interventions can be specified with a list containing a pair of functions: an "intervention" function which outputs desired treatment assignments and a "g.star" function which outputs desired treatment probabilities. These functions can take treatment and covariates as arguments and must produce treatment assignments and probabilities respectively, each with the same dimensions as the observed treatment. The function \code{makeITT()} creates list of functions corresponding to the binary treat-all and treat-none static interventions, which can be used as a template for specifying more complex interventions.

When specifying dynamic interventions using covariate names, it is important to set \code{RenameCovs = FALSE}, as otherwise \pkg{concrete} may rename covariates in the process of one-hot encoding categorical variables. "intervention" functions should take three inputs with the first being a data.table containing columns observed treatment values (here \code{ObservedTrt}), the second being a data.table of baseline covariates (here \code{Covariates}), and third being a data.table of propensity scores for the observed treatment values (here \code{PropScore}). The output of the "intervention" function should be a data.table with the same dimensions and names as the input observed treatment data.table, but containing the intervention treatment values instead. "g.star" functions take an additional fourth argument which should be a data.table of intervention treatment values (i.e. the output of the "intervention" function) and returns a data.table with one column containing the intervention propensity scores for each subject. Below we present an example of specifying dynamic interventions based on subjects' age being greater than 60 or not, captured by the "age" column.

\begin{verbatim}
TreatOver60 <- list(
    "intervention" = function(ObservedTrt, Covariates, PropScore) {
        # make an output data.table with the same dimensions as observed treatment 
        Intervened <- data.table::copy(ObservedTrt)
        
        # generalized to handle multiple treatment columns, all treatment
        #   columns are assigned 1 if for rows where age is >60
        Intervened[, (colnames(ObservedTrt)) := lapply(.SD, function(a) {
            as.numeric(Covariates[["age"]] > 60)
        }), .SDcols = colnames(ObservedTrt)]
        return(Intervened)
    }, 
    "g.star" = function(Treatment, Covariates, PropScore, Intervened) {
        # Probability set to 1 if an individual's observed treatment, "Treatment",  
        #   equals their assigned treatment, "Intervened"
        Probability <- data.table::data.table(1 * sapply(1:nrow(Treatment), function(i) 
            all(Treatment[i, ] == Intervened[i, ])))
        return(Probability)
    }
)
TreatUnder60 <- list(
    "intervention" = function(ObservedTrt, Covariates, PropScore) {
        Intervened <- data.table::copy(ObservedTrt)
        Intervened[, (colnames(ObservedTrt)) := lapply(.SD, function(a) {
            as.numeric(Covariates[["age"]] <= 60)
        }), .SDcols = colnames(ObservedTrt)]
        return(Intervened)
    }
    # if a g.star function is not specified, the makeITT() g.star function, 
    #   i.e. the g.star function above, will be used.
)

ConcreteArgs <- formatArguments(
    DataTable = data, 
    EventTime = "time", 
    EventType = "status", 
    Treatment = "trt",
    Intervention = list("Treat>60" = TreatOver60, 
                        "Treat<=60" = TreatUnder60), 
    TargetTime = 365.25/2 * (6:12), 
    TargetEvent = 1:2, 
    CVArg = list(V = 10), 
    RenameCovs = FALSE, ## turn off covariate pre-processing ##
    Verbose = FALSE
)
\end{verbatim}

The \code{TargetEvent} argument specifies the event types of interest. Event types must be be coded as integers, with non-negative integers reserved for censoring. If \code{TargetEvent} is left \code{NULL}, then all positive integer event types in the observed data will be jointly targeted. In the \code{pbc} dataset, there are 3 event values encoded by the\code{status} column: 0 for censored, 1 for transplant, and 2 for death. To analyze \code{pbc} with transplants treated as right censoring, \code{TargetEvent} should be set to 2, whereas for a competing risks analysis one could either leave \code{TargetEvent = NULL} or set \code{TargetEvent = 1:2} as in the above example.

The \code{TargetTime} argument specifies the times at which the cause-specific absolute risks or event-free survival are estimated. Target times should be restricted to the time range in which target events are observed and \code{formatArguments()} will return an error if target time is after the last observed failure event time. If no \code{TargetTime} is provided, then \pkg{concrete} will target the last observed event time, though this is likely to result in a highly variable estimate if prior censoring is substantial. The \code{TargetTime} argument can either be a single number or a vector, as one-step TMLE can target cause-specific risks at multiple times simultaneously. 

\subsubsection{Estimator specification} \label{EstimationSpec}
The \code{formatArguments()} function can be used to modify the estimation procedure. The arguments of \code{formatArguments()} involved in estimation are the cross-validation setup \code{CVArg}, the Superlearner candidate libraries \code{Model}, the software backends \code{PropScoreBackend} and \code{HazEstBackend}, and the practical TMLE implementation choices \code{MaxUpdateIter}, \code{OneStepEps}, and \code{MinNuisance}. Note that \code{Model} is used in this section in line with common usage in statistical software, rather than to refer to formal statistical or causal models as in preceding sections. 

Cross-validation is implemented using \code{origami::make\_folds()} and using the input of the \code{CVArg} argument. If no input is provided into \code{CVArg}, the default cross-validation setup follows the recommendations in \citep{phillips_practical_2023}. Cross-validation folds are stratified by event type and the number of folds ranges from 2 for datasets with greater than 10000 independent observations to LOOCV for datasets with fewer than 30 independent observations. Chapter 5 of the online Targeted Learning Handbook \citep{malenica_chapter_nodate} demonstrates the specification of several other cross-validation schemes.

Super Learner libraries for estimating nuisance parameters are specified through the \code{Model} argument. The input should be a named list with an element for the treatment variable and one for each event type including censoring as illustrated in the following code example. The list element corresponding to treatment must be named with the column name of the treatment variable, and the list elements corresponding to each event type must be named by the character which corresponds to the numeric value of the event type (e.g. "0" for censoring). Any missing specifications will be filled in with defaults, and the resulting list of libraries can be accessed in the output \code{.[["Model"]]} and further edited by the user, as shown below.

\begin{verbatim}
# specify regression models
ConcreteArgs$Model <- list(
    "trt" = c("SL.glmnet", "SL.bayesglm", "SL.xgboost", "SL.glm", "SL.ranger"),
    "0" = NULL, # will use the default library
    "1" = list(Surv(time, status == 1) ~ trt, Surv(time, status == 1) ~ .),
    "2" = list("Surv(time, status == 2) ~ trt", "Surv(time, status == 2) ~ .")
)
\end{verbatim}

In \pkg{concrete}, propensity scores are by default estimated using the with candidate algorithms \code{c("xgboost", "glmnet")} implemented by packages \CRANpkg{xgboost} \citep{chen_xgboost_2022} and \CRANpkg{glmnet} \citep{friedman_regularization_2010}. For further details about these packages, see their respective package documentations.

For estimating the necessary conditional hazards, \pkg{concrete} currently relies on a discrete Superlearner consisting of a library of Cox models implemented by \code{survival::coxph()} evaluated on cross-validated partial-likelihood loss as detailed in Section \ref{haz-est}. Support for estimation of hazards using coxnet \citep{simon_regularization_2011}, Poisson-HAL and other methods is planned in future package versions. The default Cox specifications are a treatment-only regression and a main-terms regression including treatment and all covariates. These models can be specified as strings or formulas as can be seen in the above example.

As detailed by Eq. \eqref{one-step} and \eqref{one-step-stop}, the one-step TMLE update step involves recursively updating cause-specific hazards, summing along small steps, scaled by a multiplicative factor \(\epsilon_b\). The default initial scaling factor is 0.1, and each time an update step would not decrease the mean estimated EIC, the step size scaling factor is halved and the update step is re-tried. The \code{MaxUpdateIter} argument is used to provide a definite stop to the recursive TMLE update. The default of 500 steps should be sufficient for most applications, but may need to be increased when targeting estimands with many components or for rare events. The presence of practical positivity sparsity can also result in slow TMLE convergence, but increasing \code{MaxUpdateIter} would not be not an adequate solution there as the resulting TMLE estimates and inference may still be unreliable.
The \code{MinNuisance} argument can be used to specify a lower bound for the product of the propensity score and lagged survival probability for remaining uncensored; this term is present in the denominator of the efficient influence function and enforcing a lower bound decreases estimator variance at the cost of introducing bias but improving stability.

\subsubsection{ConcreteArgs objects}
\label{concreteargs}
The \code{"ConcreteArgs"} output of \code{formatArguments()} is an environment containing the estimation specification as objects that can be modified by the user. The modified \code{"ConcreteArgs"} object should then be passed back through \code{formatArguments()} to check the modified estimation specification. 

\begin{verbatim}
# decrease the maximum tmle update number to 50
ConcreteArgs$MaxUpdateIter <- 50

# add a candidate regression with treatment interactions
ConcreteArgs[["Model"]][["2"]][[3]] <- "Surv(time, status == 2) ~ trt*."

# validate new estimation specification
ConcreteArgs <- formatArguments(ConcreteArgs)
\end{verbatim}

\code{"ConcreteArgs"} objects can be printed to display summary information about the specified estimation problem,

\begin{verbatim}
print(ConcreteArgs, Verbose = FALSE)
\end{verbatim}

Below, we can see that the specified analysis is for two competing risks (Target Events: 1, 2) under interventions "A=1" and "A=0" assigning all subjects to treated and control arms, respectively. Objects in the \code{"ConcreteArgs"} environment can be interrogated directly for details about any particular aspect of the estimation specification. For instance, as mentioned before the one-hot encoding of covariates can be seen in a table by \code{attr(.[["DataTable"]], "CovNames")}, intervention treatment assignments can be checked at \code{.[["Regime]]}.

\begin{figure}[H]
\includegraphics[width=\linewidth]{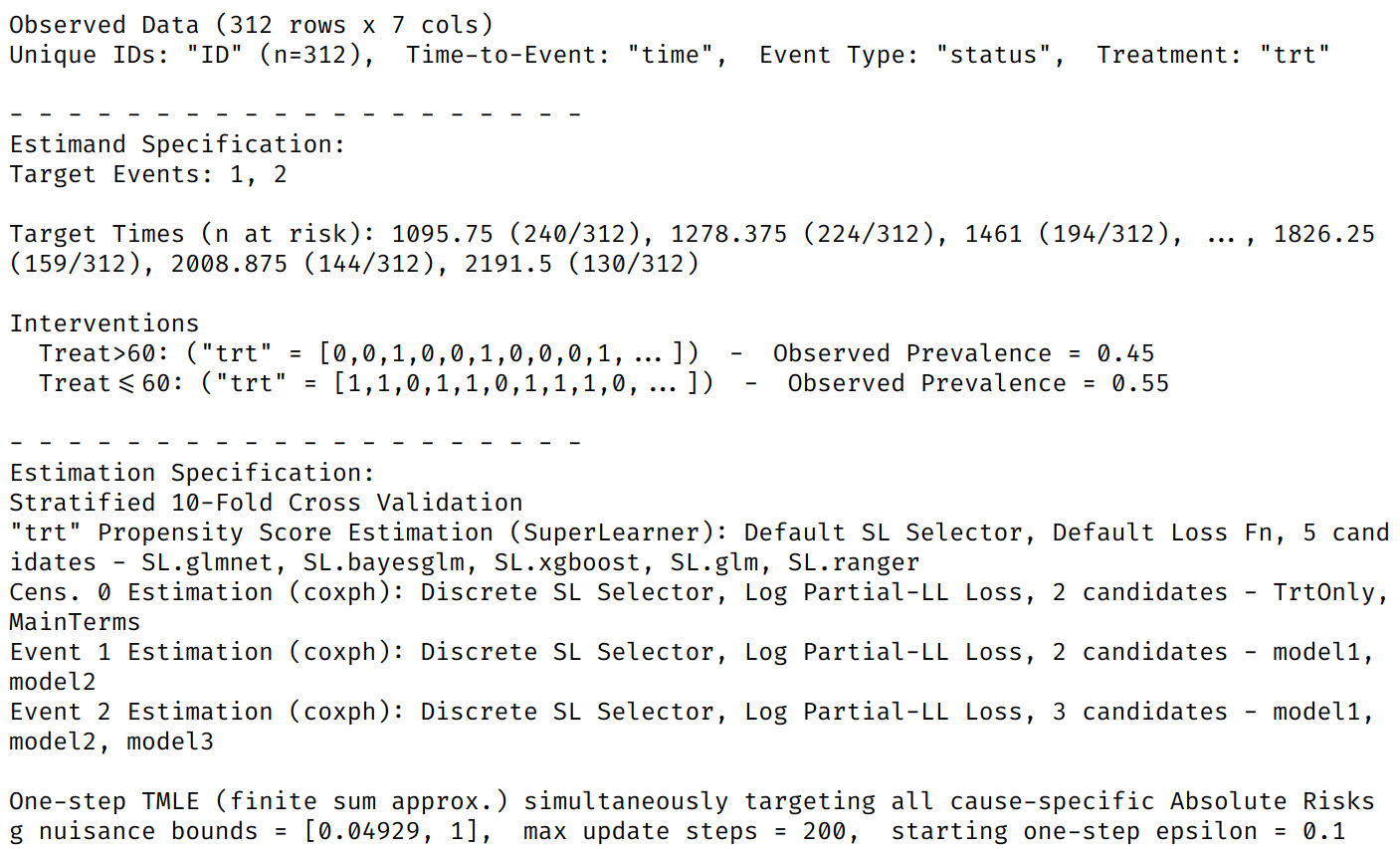}
\end{figure}

\subsection{Estimation}
\label{doConcrete}
Adequately specified \code{"ConcreteArgs"} objects can then be passed into the \code{doConcrete()} function which will then perform the specified TMLE analysis. The output is an object of class \code{"ConcreteEst"} which contains TMLE point estimates and corresponding estimated influence curves for the cause-specific absolute risks for each targeted event at each targeted time under each intervention. If the \code{GComp} argument is set to \code{TRUE}, then a Super Learner-based g-formula plugin estimate of the targeted risks will be included in the output. 

\begin{verbatim}
ConcreteEst <- doConcrete(ConcreteArgs)
\end{verbatim}

We have reviewed the one-step continuous-time TMLE implementation in Section \ref{estimation}, so here we will name the non-exported functions in \code{doConcrete()} which perform each of the steps of the one-step continuous-time survival TMLE procedure, in case users wish to explore the implementation in depth.

The cross-validation (Section \ref{cv}) is checked and evaluated in \code{formatArguments()}, returning fold assignments as the \code{.[["CVFolds"]]} element of the \code{"ConcreteArgs"} object.

The initial estimation of nuisance parameters and is performed by the function \code{getInitialEstimate()} which depends on \code{getPropScore()} for propensity scores (Section \ref{trtps-est}) and \code{getHazEstimate()} for the conditional hazards (Section \ref{haz-est}).

Computing of EICs is done by \code{getEIC()} which is used within the \code{doTmleUpdate()} function which performs the one-step TMLE update procedure (Section \ref{EIC}).

\subsubsection{ConcreteEst objects}
\label{concreteest}
The print method for \code{"ConcreteEst"} objects summarizes the estimation target and displays diagnostic information about TMLE update convergence, intervention-related nuisance parameter bounding, and the nuisance parameter Super Learners.

\begin{verbatim}
    print(ConcreteEst, Verbose = FALSE)
\end{verbatim}
\begin{figure}[H]
\center
\includegraphics[width=\linewidth]{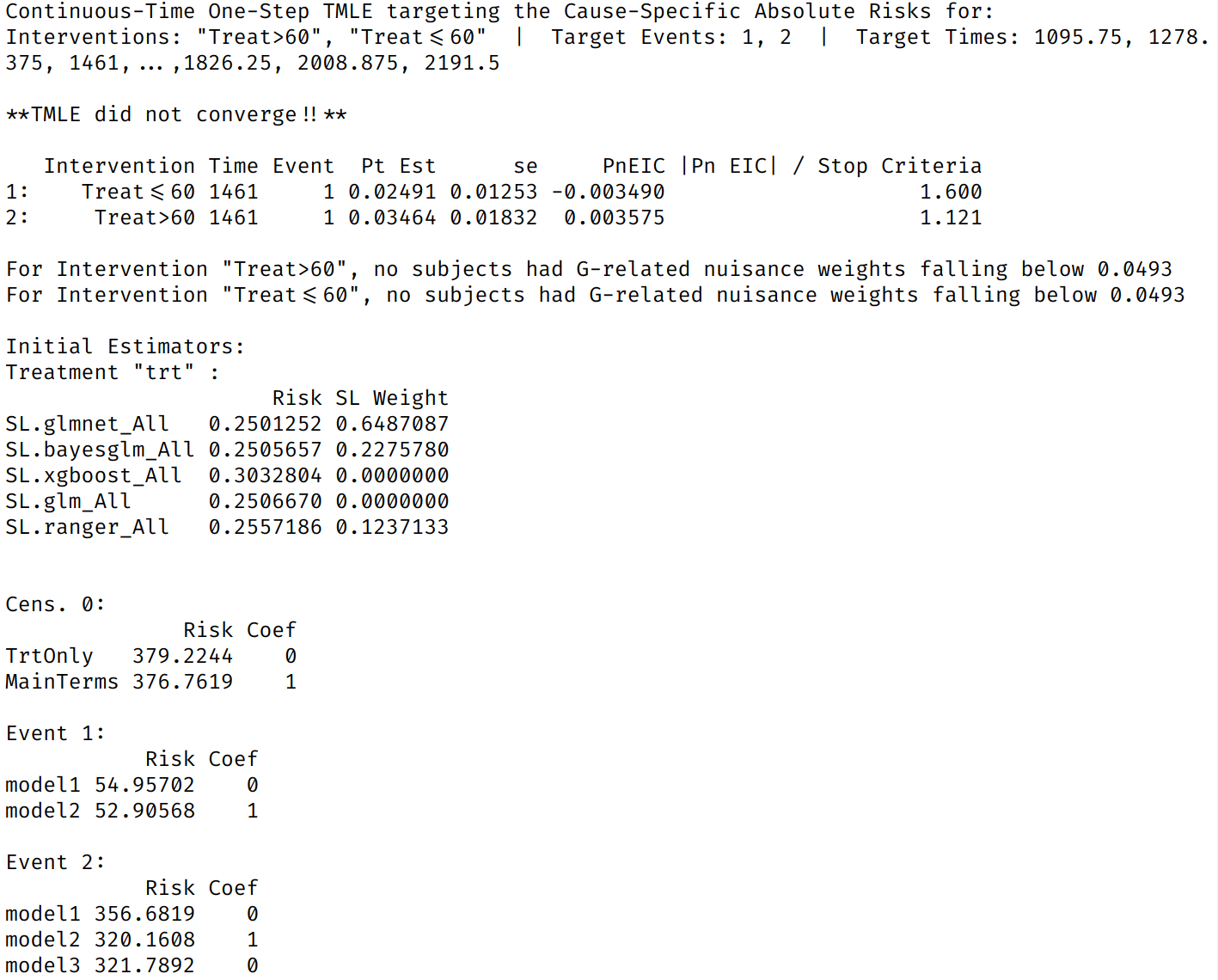}
\end{figure}

If TMLE has not converged, the mean EICs that have not attained the desired cutoff, i.e. Eq \eqref{one-step}, will be displayed in a table. For instance, we can see above that the absolute value of the mean EIC for intervention \samp{Treat>60} at time 1461 for event 1 has not reached the stopping criteria and is $1.6$ times larger than the stopping criteria. Increasing the the maximum number of TMLE update iterations via \code{MaxUpdateIter} can allow TMLE to finish updating nuisance parameters, though at target time points when few events have yet occurred even small mean EIC values may not meet the convergence criteria and adequate convergence may require many iterations. 

The extent to which the intervention-related nuisance parameters (i.e. propensity scores and probabilities of remaining uncensored) have been lower-bounded is also reported for each intervention both in terms of the percentage of nuisance weights that have been bounded and the percentage of subjects with bounded nuisance weights. If users suspect possible positivity issues, the plot method for \code{"ConcreteEst"} objects can be used to visualize the distribution of estimated propensity scores for each intervention, with the red vertical line marking the cutoff for lower-bounding.

\begin{verbatim}
    plot(ConcreteEst, ask  = FALSE)
\end{verbatim}

\begin{figure}[H]
\includegraphics[width=\linewidth]{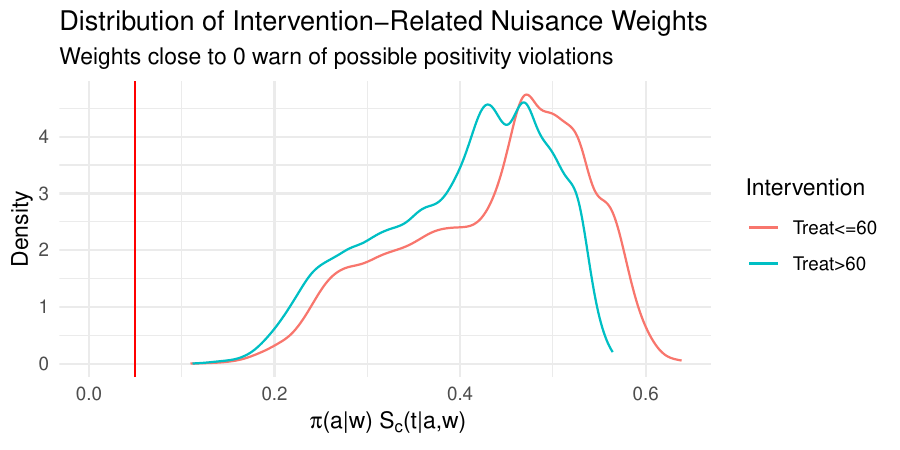}
\caption{}
\label{fig:ps-diag}
\end{figure}

Intervention-related nuisance parameters with values close to 0 indicate the possibility of positivity violations and may warrant re-examining the target time(s), interventions, and covariate adjustment sets. In typical survival applications, positivity issues may arise when targeting times at which some subjects are highly likely to have been censored, or if certain subjects are unlikely to have received a desired treatment intervention. As positivity violations not only impact causal interpretability, but also estimator behaviour, we urge users to re-consider their target analyses; \citep{petersen_diagnosing_2012} provides guidance on reacting to positivity issues.

The last three tables ("Cens. 0", "Event 1", and "Event 2") in the above code output show the candidate estimators of nuisance parameters, summarized with the cross-validated risk of each candidate estimator followed by their weight in the corresponding Super Learners. 

\subsection{Producing outputs}
\label{getoutput}
\code{getOutput()} takes as an argument the \code{"ConcreteEst"} object returned by \code{doConcrete()} and can be used to produce tables and plots of the cause-specific risks, risk differences, and relative risks. By default \code{getOutput()} returns a \code{data.table} with point estimates and pointwise standard errors for cause-specific absolute risks, risk differences, and risk ratios. By default, the first listed intervention is used as the "treated" group while the second is considered "control"; other contrasts can be specified via the \code{Intervention} argument. Below we show a subset of the relative risk estimates produced by the "nutshell" estimation specification for the pbc dataset. 

\begin{verbatim}
ConcreteOut <- getOutput(ConcreteEst = ConcreteEst, 
                         Estimand = "RD",
                         Intervention = 1:2, 
                         GComp = TRUE, 
                         Simultaneous = TRUE, 
                         Signif = 0.05)
head(ConcreteOut, 12)
\end{verbatim}

\begin{verbatim}
#>        Time Event  Estimand             Intervention Estimator   Pt Est    se
#> 1: 1095.750     1 Risk Diff [Treat>60] - [Treat<=60]      tmle  0.00800 0.018
#> 2: 1095.750     1 Risk Diff [Treat>60] - [Treat<=60]     gcomp  0.00300    NA
#> 3: 1095.750     2 Risk Diff [Treat>60] - [Treat<=60]      tmle -0.02000 0.040
#> 4: 1095.750     2 Risk Diff [Treat>60] - [Treat<=60]     gcomp  0.00025    NA
#> 5: 1278.375     1 Risk Diff [Treat>60] - [Treat<=60]      tmle  0.00800 0.018
#> 6: 1278.375     1 Risk Diff [Treat>60] - [Treat<=60]     gcomp  0.00300    NA
#>    CI Low CI Hi SimCI Low SimCI Hi
#> 1: -0.027 0.043    -0.039    0.055
#> 2:     NA    NA        NA       NA
#> 3: -0.099 0.058    -0.130    0.086
#> 4:     NA    NA        NA       NA
#> 5: -0.027 0.043    -0.039    0.055
#> 6:     NA    NA        NA       NA
\end{verbatim}

\begin{figure}[h]
    \centering
    \includegraphics[width = \linewidth]{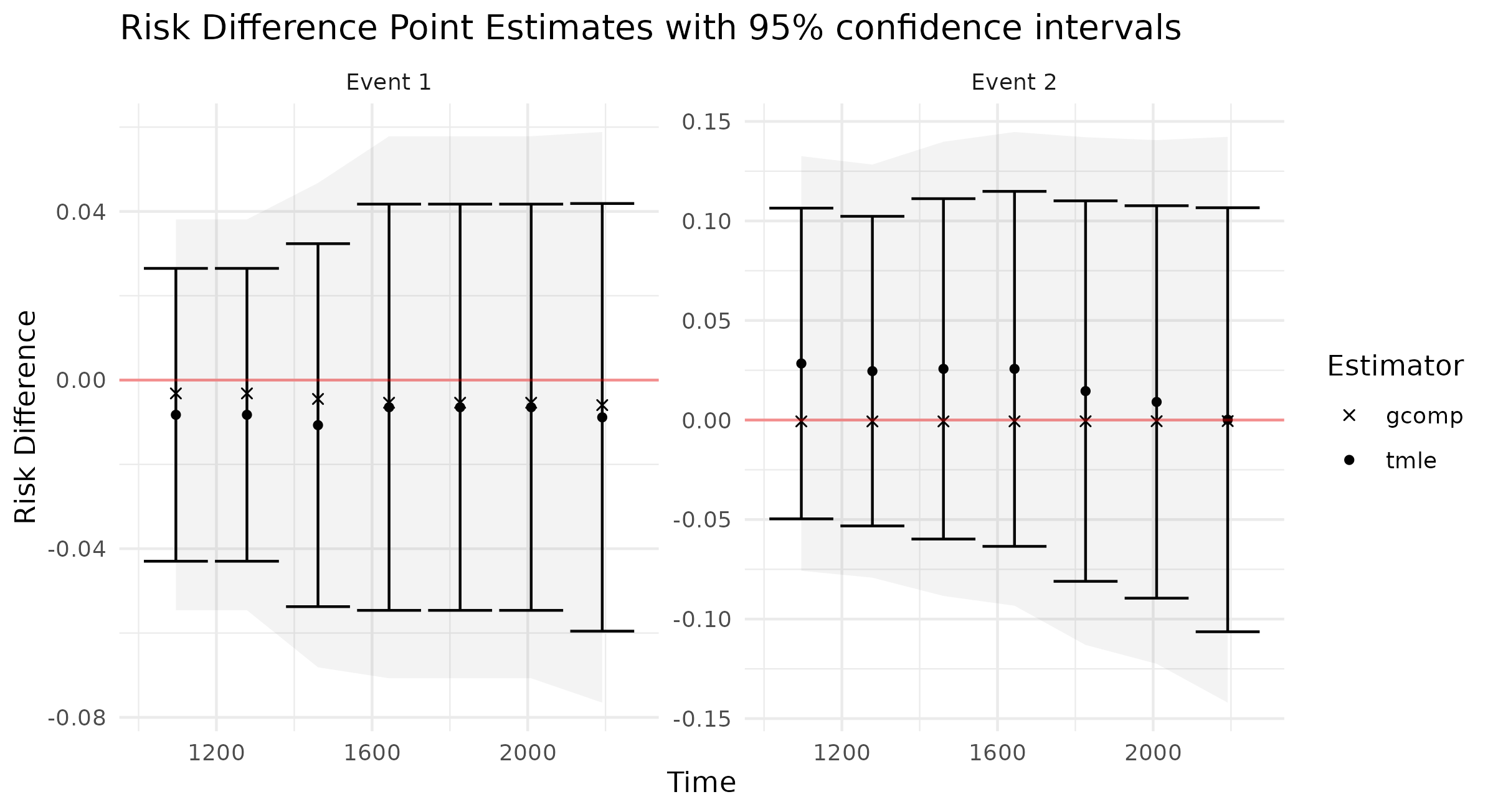}
    \caption{Estimated effect of D-penicillamine on the differences in cumulative incidence of the competing risks of transplant (Event 1) and death (Event 2) in the PBC dataset. Simultaneous 95\% confidence bands are shown as gray ribbons. Error bars plot TMLE's point-wise 95\% confidence intervals. Null treatment effects are plotted as horizontal red lines.}
    \label{fig:nutshell-pbc-rd}
\end{figure}

From left to right, the first five columns describe the estimands (target times, target events, estimands, and interventions) and estimators. The subsequent columns show the point estimates, estimated standard error, confidence intervals and simultaneous confidence bands. The \code{Signif} argument (set to a default of 0.05) specifies the desired double-sided alpha which is then used to compute confidence intervals, and the \code{Simultaneous} argument specifies whether or not to compute a simultaneous confidence band for all output TMLE estimates. Here we also see that when estimands involve many time points or multiple events, tables may be difficult to interpret at a glance. Instead plotting can make treatment effects and trends more visually interpretable, as was shown in Figure \ref{fig:nutshell-pbc-rd}.

The \code{plot} method for \code{"ConcreteOut"} object invisibly returns a list of \code{"ggplot"} objects, which can be useful for personalizing graphical outputs. Importantly, users should note that plots do not currently indicate if TMLE has converged or if positivity may be an issue; users must therefore take care to examine the diagnostic output of the \code{"ConcreteEst"} object prior to producing effect estimates using \code{getOutput()}.

\begin{verbatim}
library(concrete)
data <- survival::pbc[, c("time", "status", "trt", "age", "sex", "albumin")]
data <- subset(data, subset = !is.na(data$trt))
data$trt <- data$trt - 1

# Specify Analysis
ConcreteArgs <- formatArguments(
    DataTable = data, 
    EventTime = "time",
    EventType = "status", 
    Treatment = "trt",
    Intervention = 0:1, 
    TargetTime = 365.25/2 * (6:12), 
    TargetEvent = 1:2, 
    CVArg = list(V = 10), 
    Verbose = FALSE
)

# Compute
ConcreteEst <- doConcrete(ConcreteArgs)

# Return Output
ConcreteOut <- getOutput(ConcreteEst, Estimand = "RD", Simultaneous = TRUE)
plot(ConcreteOut, NullLine = TRUE, ask = FALSE)
\end{verbatim}

\subsubsection{Specifying dynamic interventions}
\begin{verbatim}
TreatOver60 <- list(
    "intervention" = function(ObservedTrt, Covariates, PropScore) {
        # make an output data.table with the same dimensions as observed treatment 
        Intervened <- data.table::copy(ObservedTrt)
        
        # generalized to handle multiple treatment columns, all treatment
        #   columns are assigned 1 if for rows where age is >60
        Intervened[, (colnames(ObservedTrt)) := lapply(.SD, function(a) {
            as.numeric(Covariates[["age"]] > 60)
        }), .SDcols = colnames(ObservedTrt)]
        return(Intervened)
    }, 
    "g.star" = function(Treatment, Covariates, PropScore, Intervened) {
        # Probability set to 1 if an individual's observed treatment, "Treatment",  
        #   equals their assigned treatment, "Intervened"
        Probability <- data.table::data.table(1 * sapply(1:nrow(Treatment), function(i) 
            all(Treatment[i, ] == Intervened[i, ])))
        return(Probability)
    }
)
TreatUnder60 <- list(
    "intervention" = function(ObservedTrt, Covariates, PropScore) {
        Intervened <- data.table::copy(ObservedTrt)
        Intervened[, (colnames(ObservedTrt)) := lapply(.SD, function(a) {
            as.numeric(Covariates[["age"]] <= 60)
        }), .SDcols = colnames(ObservedTrt)]
        return(Intervened)
    }
    # if a g.star function is not specified, the makeITT() g.star function, 
    #   i.e. the g.star function above, will be used.
)

ConcreteArgs <- formatArguments(
    DataTable = data, 
    EventTime = "time", 
    EventType = "status", 
    Treatment = "trt",
    Intervention = list("Treat>60" = TreatOver60, 
                        "Treat<=60" = TreatUnder60), 
    TargetTime = 365.25/2 * (6:12), 
    TargetEvent = 1:2, 
    CVArg = list(V = 10), 
    RenameCovs = FALSE, ## turn off covariate pre-processing ##
    Verbose = FALSE
)
\end{verbatim}

\subsection{Summary}
\label{summary}
This paper introduces the \pkg{concrete} R package implementation of continuous-time estimation for absolute risks of right censored time-to-event outcomes. The package fits into the principled causal-inference workflow laid out by the targeted learning roadmap and allows fully compatible estimation of cause-specific absolute risk estimands for multiple events and at multiple times. The \code{formatArguments()} function is used to specify desired analyses, \code{doConcrete()} performs the specified analysis, and \code{getOutput()} is used to produce formatted output of the target estimands. Cause-specific hazards can be estimated using ensembles of proportional hazards regressions and flexible options are available for estimating treatment propensities. Confidence intervals and confidence bands can be computed for TMLEs, relying on the asymptotic linearity of the TMLEs. We are currently looking into adding support for estimating cause-specific risks using coxnet and HAL-based regressions, as well as supporting stochastic interventions with multinomial or continuous treatment variables. 

\section{Acknowledgements}
This work was completed as a part of The Joint Initiative for Causal Inference funded through a philanthropic donation from Novo Nordisk.

\bibliography{RJreferences}

\address{David Chen\\
  Department of Biostatistics, University of California, Berkeley\\
  2121 Berkeley Way, Berkeley, CA 94720\\
  USA\\
  \email{david.chen49@berkeley.edu}}

\address{Helene C. W. Rytgaard\\
    Section of Biostatistics, Department of Public Health, University of Copenhagen\\
    Øster Farimagsgade 5, 1014 Copenhagen\\
    Denmark\\
    \email{hely@sund.ku.dk}}

\address{Jens M. Tarp\\
  Novo Nordisk A/S\\
  Vandtårnsvej 114, DK-2860 Søborg\\
  Denmark\\
  \email{author1@work}}

\address{Edwin C. H. FONG\\
  Department of Statistics and Actuarial Science\\
  Rm 230, Run Run Shaw Building, Pokfulam Road\\
    Hong Kong\\
  \email{chefong@hku.hk}}

\address{Maya L. Petersen\\
  Department of Biostatistics, University of California, Berkeley\\
  2121 Berkeley Way, Berkeley, CA 94720\\
  USA\\
  \email{mayaliv@berkeley.edu}}

\address{Mark J. van der Laan\\
  Department of Biostatistics, University of California, Berkeley\\
  2121 Berkeley Way, Berkeley, CA 94720\\
  USA\\
  \email{laan@stat.berkeley.edu}}

\address{Thomas A. Gerds\\
    Section of Biostatistics, Department of Public Health, University of Copenhagen\\
    Øster Farimagsgade 5, 1014 Copenhagen\\
    Denmark\\
    \email{tag@biostat.ku.dk}}